\documentclass[prb,twocolumn,notitlepage,superscriptaddress,showpacs,amsmath,amssymb]{revtex4-1}
\usepackage{hyperref,graphicx}

\begin{document}

 \title{Dynamical transport measurement of the Luttinger parameter in helical edges states of 2D topological insulators}

\author{Tobias M\"uller}
\email{tobias.mueller@physik.uni-wuerzburg.de}
\affiliation{Institute for Theoretical Physics and Astrophysics, University of W\"urzburg, D-97074 W\"urzburg, Germany}
\author{Ronny Thomale}
\affiliation{Institute for Theoretical Physics and Astrophysics, University of W\"urzburg, D-97074 W\"urzburg, Germany}
\author{Bj\"orn Trauzettel}
\affiliation{Institute for Theoretical Physics and Astrophysics, University of W\"urzburg, D-97074 W\"urzburg, Germany}
\author{Erwann Bocquillon}
\affiliation{Laboratoire Pierre Aigrain, Ecole Normale Sup\'erieure-PSL Research University, CNRS, Universit\'e Pierre et Marie Curie-Sorbonne
Universit\'es, Universit\'e Paris Diderot-Sorbonne Paris Cit\'e, 24 rue Lhomond, 75231 Paris Cedex 05, France}
\affiliation{EP3, University of W\"urzburg, D-97074 W\"urzburg, Germany}
\author{Oleksiy Kashuba}
\affiliation{Institute for Theoretical Physics and Astrophysics, University of W\"urzburg, D-97074 W\"urzburg, Germany}

\date{\today}

\begin{abstract}
One-dimensional (1D) electron systems in the presence of Coulomb interaction are described by Luttinger liquid theory. The strength of Coulomb interaction in the Luttinger liquid, as parameterized by the Luttinger parameter $K$, is in general difficult to measure. This is because $K$ is usually hidden in powerlaw dependencies of observables as a function of temperature or applied bias. We propose a dynamical way to measure $K$ on the basis of an electronic time-of-flight experiment. We argue that the helical Luttinger liquid at the edge of a 2D topological insulator constitutes a preeminently suited realization of a 1D system to test our proposal. This is based on the robustness of helical liquids against elastic backscattering in the presence of time reversal symmetry.
\end{abstract}


\maketitle

\section{Introduction}

Since the motion of electrons is strongly geometrically constrained in 1D conductors, Coulomb interactions have particularly pronounced effects on transport properties. The paradigm of quasi-free quasiparticles (as in the Fermi liquid), valid in higher dimensions, is then replaced by a collective description of electronic excitations in terms of bosonic density waves in the Luttinger liquid picture. \cite{Giamarchi2003}
Due to their capacitive nature, Coulomb interactions can be conveniently evidenced by radio frequency (RF) measurements, as exemplified in the chiral edge states of the quantum Hall effect. \cite{Sukhodub2004,Kumada2011, Bocquillon2013b,Kamata2014}
Similarly Quantum spin Hall (QSH) insulators exhibit transport behavior, which is governed by a pair of counterpropagating helical edge states. The right and left movers at a given boundary of the physical system carry opposite spin, \cite{Wu2006,Xu2006} rendering them robust against elastic backscattering. So far, experimental efforts have largely focused on the prospects for topologically protected edge state transport \cite{Koenig2007,Roth2009,Knez2011,jorgo} or topological Josephson junctions. \cite{Fu2009,Crepin2014,Bocquillon2016,Deacon2016} QSH insulators, however, also enable studies of Coulomb interaction in one-dimensional conductors. \cite{Li2015} The interplay of Coulomb interaction and randomness in spin-orbit coupling, e.g. due to rough edges, creates an additional source of inelastic backscattering in helical liquids. \cite{Strom2010,Schmidt2012,Crepin2012,Geissler2014,Kainaris2014} Still, these corrections to transport are usually suppressed at low energy scales. Hence, unlike most other 1D conductors such as carbon nanotubes or semiconducting nanowires, mobilities in helical Luttinger liquids (hLLs) can be high, and mean-free paths exceeding $l\sim10\,\mu\text{m}$ have already been observed. \cite{Bendias2017} As further realizations of QSH insulators with large bulk gaps start to enter the stage~\cite{jorgo} and step edges on high symmetry surfaces of topological crystalline insulators likewise give rise to novel scenarios of 1D helical edge channels,\cite{sessi} this pushes the boundaries within which coherent 1D transport allows to be investigated, and calls for new studies of dynamical transport inspired by their analogue in the quantum Hall effect. \cite{Hofer2013,Inhofer2013,Garate2012}

\begin{figure}
\centering
\includegraphics[width=\columnwidth]{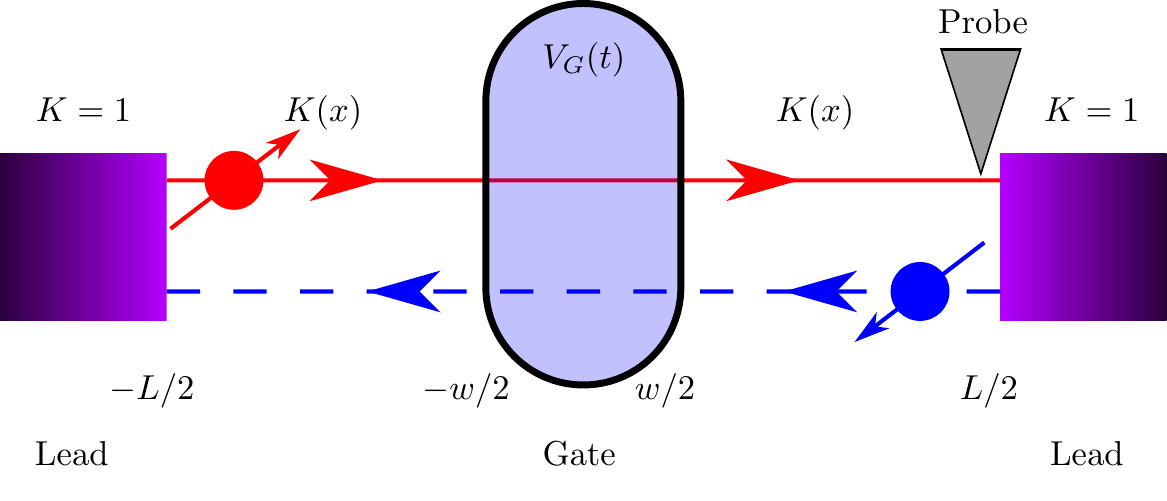}
\caption{Geometry proposed for measuring the interaction parameter K of a one-dimensional conducting channel of length $L$ connected to two Fermi leads reservoirs with $K=1$. A gate voltage $V_G(t)$ is applied via a gate of length $w$ in the middle of the channel. The signal is either read out as a current via one of the leads or capacitively via a probe at the contact.}
\label{fig:setup}
\end{figure}

In this article, we investigate the general problem of inhomogeneous Luttinger liquids, in which the interaction parameter $K(x)$ is assumed to depend on the position coordinate $x$ in the channel.\cite{Tobi}
The dynamics of a Luttinger liquid can be characterized by the dispersion of the excitations and interactions inside or between the transport channels. In the particular case of hLLs, it can be described by the renormalized Fermi velocity $u$ and the Luttinger parameter $K$. The dynamics of homogeneous hLLs have recently been investigated in several theoretical works, \cite{Dolcetto2015,Calzona2015,Calzona2016} with a focus on effects such as spin-charge separation.
Inhomogeneous Luttinger liquids have first been studied in the context of investigating the $K$ dependence of the Landauer conductivity, which was eventually concluded to be independent of $K$.\cite{Maslov1995,Pono1995,Safi1995} Indeed, assuming a continuous spatial dependence of $K$ at the scale of the Fermi wavelength, \cite{Thomale2011} the conductance between left and right leads is solely determined by the contacts, which, being effectively non-interacting higher dimensional Fermi liquids, feature $K=1$. While reflections do occur due to variations of $K$, the DC amplitudes sum up to the quantized conductance expected in the absence of interactions.\cite{Safi1995,Safi1999} The AC regime was shown to in principle inherit a $K$ dependence from the spatial interaction profile in the wire.\cite{Thomale2011} 
As a consequence, in DC experiments, the Luttinger parameter does not surface in the conductance value itself, and instead would have to be tediously extracted as an exponent from the dependence of the tunneling density of states $\nu(E) \propto |E|^{(1-K)^{2}/2K}$ on energy\cite{Luther1974} or the temperature dependence of the conductance $G(T) \propto T^{2(1-K)/K}$ through a tunnel barrier.\cite{Kane1992} An alternative possibility to determine the Luttinger parameter has been identified through the finite frequency current noise,\cite{Trauzettel2004,Dolcini2005} but is likewise rather demanding to perform experimentally.

We propose a new approach to measure the Luttinger parameter $K$ in 1D conductors in the geometry presented in Fig.~\ref{fig:setup}. Using top-gate electrodes, we define regions in which interactions are efficiently screened with $K\simeq 1$ over a scale of a few tens of nanometers, much smaller than $l$. Unscreened regions conserve an interaction parameter $K\neq1$. Another narrower top-gate electrode is capacitively coupled to the conducting channel to generate an AC excitation in the channel. We show, that the AC conductance between the excitation gate and a readout contact (ohmic or capacitive) exhibits a simple $K$-dependence. Applying a time-dependent voltage to the gate, as shown in Fig.~\ref{fig:setup}, we create electron- and hole-like excitations, which scatter on the spatial variations of $K(x)$. The scattered charge pulses can be detected via time-resolved measurements with realistic sub-nanosecond resolution \cite{Kamata2014} in the readout lead. Alternatively, we calculate the finite frequency admittance $g(\omega)$, and demonstrate that it allows for a reliable extraction of $K$ in the edge channels. Both methods rely on demonstrated microwave techniques, making this proposal experimentally feasible. Furthermore, while QSH insulators are a prototypical example to test our proposal, it is not limited to QSH edge states, but applies to any (non-chiral) Luttinger liquid. For QSH insulators, we take advantage of the large mean free-path in the QSH edge states, the absence of backscattering at the center gate voltage, and the easiness to design on-demand geometries. In terms of theoretical methodology, we employ an equation of motion perspective as established in Refs.~\onlinecite{Maslov1995,Thomale2011}, which proves ideally suited for this task. 

The article is organized as follows. In Section~\ref{sec_iLL}, we recap the relevant properties of the inhomogeneous Luttinger liquid for our purposes. Subsequently, in Section~\ref{sec_time}, we describe the time-resolved transport in the setup shown in Fig.~\ref{fig:setup}. In further sections, we go from time to frequency domain (Section~\ref{SectionFreq}) and discuss the influence of an arbitrary profile of $K(x)$ on our results (Section~\ref{sec_K}) before we summarize our results in Section~\ref{sec_sum}. Some technical details of the calculations are delegated to the Appendix.

\section{Inhomogeneous Luttinger liquid}
\label{sec_iLL}

Consider a one-dimensional interacting helical edge channel.
The bosonized Hamiltonian of the system with space dependent interaction parameter $K(x)$ and mode velocity $u(x)$, including the coupling to a gate potential $\varphi(x)$, introduced by a minimal coupling term, reads:
\begin{multline}
H = \frac{1}{2 \pi} \int dx\, u(x)
\left[ K(x) \bigl(\pi\Pi(x)\bigr)^2 + \frac{\bigl(\partial_x\Phi(x)\bigr)^2}{K(x)} \right]
\\
-\frac{e}{\pi} \int dx \, \varphi(x) \partial_x \Phi(x).
\label{eq:hamiltonian}
\end{multline}
The mode velocity $u(x)$ and Luttinger parameter $K(x)$ are determined by the local momentum preserving electron-electron interactions inside the 1D region~\cite{Giamarchi2003} ( in this section we work in units where $\hbar = 1$, for simplicity, but will restore physical units in subsequent sections).
Since the gate is capacitively coupled to the conducting channel, no charge transfer is possible between gate and conducting channel, as reflected by the minimal coupling term. Note that this approach neglects the effect of the geometrical capacitance of the gate, which could be treated as an additional quadratic term in the Hamiltonian. \cite{Mora2010} It is, however, usually very large (against the quantum capacitance, see Sec.~\ref{SectionFreq}), and is here taken as infinite.
The values $K<1$ ($K>1$) correspond to repulsive (attractive) interactions between electrons and the non-interacting case corresponds to $K=1$ and mode velocity equal to the Fermi velocity $u=v_F$.
For Galilean invariant systems, the relation $u(x)K(x) = v_F$ holds, but we do not require this identity for our results.

The charge and the current densities can be expressed in terms of the bosonic fields as $\rho(x) = -\frac{1}{\pi} \partial_x \Phi(x)$ and $j(x) = u(x) K(x) \Pi(x)$, respectively. In order to derive the Heisenberg equations of motion for the bosonic field operator we calculate the commutators with the Hamiltonian in Eq.~\eqref{eq:hamiltonian}, resulting in~\cite{Thomale2011,Tobi}
\begin{eqnarray}
\partial_t \Phi(x) &=& \pi u(x) K(x) \Pi(x)\label{eq:phieom} \; ,\\
\partial_t \Pi(x) &=& \partial_x \left[\frac{u(x)}{K(x)} \frac{\partial_x \Phi(x)}{\pi} + \frac{e}{\pi} \varphi(x)\right] \; .
\label{eq:pieom}
\end{eqnarray}
Up to a spatial derivative, Eq.~\eqref{eq:phieom} looks like the conservation of current $\partial_t \rho(x) = -\partial_x j(x)$. Substituting Eq.~\eqref{eq:phieom} into Eq.~\eqref{eq:pieom}, we obtain an equation of motion for the field $\Phi$, which can be used as a starting point for the calculation of the response of the system to the gate voltage, 
\begin{equation}
\partial_t^2 \Phi(x) = u(x) K(x) \partial_{x} \left[\frac{u(x)}{K(x)} \partial_{x} \Phi(x) + e \varphi(x)\right].
\label{eq:eom}
\end{equation}
The boundary conditions for the equation of motion require the continuity of the field $\Phi(x)$, which corresponds to the spatial continuity of the current as expressed by Eq.~\eqref{eq:phieom}, and the continuity of the expression $\frac{u(x)}{K(x)} \partial_{x} \Phi(x) + e \varphi(x)$.
The latter condition implies the continuity of the electrochemical potential $\mu = \delta H/\delta\rho$.
Note that the equation of motion is an operator identity, as it was derived from the Heisenberg equations. Due to linearity in the field operator $\Phi(x)$ it also yields an equation for the expectation value of the operator $\langle\Phi(x)\rangle$ with respect to the ground state of the system.
Despite that all calculations are done in the frequency domain we first present real time results obtained by Fourier transform for illustration purposes.

\section{Time-resolved transport in gated helical channels}
\label{sec_time}

\begin{figure}
\centering
\includegraphics[width=.9\columnwidth,page=1]{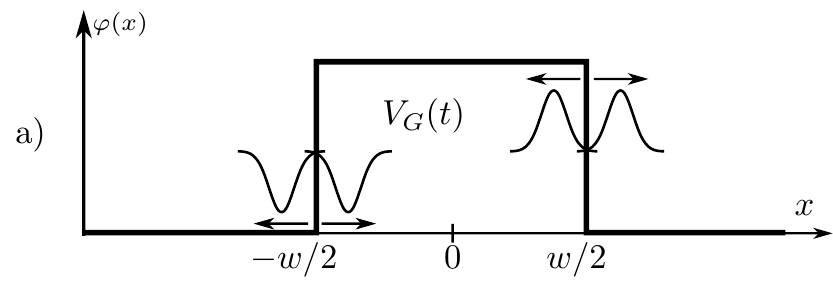}
\includegraphics[width=.9\columnwidth,page=2]{transmission.pdf}
\includegraphics[width=.9\columnwidth,page=3]{transmission.pdf}
\caption{Creation and the propagation of the excitations in a one dimensional helical channel:
%
a) The voltage pulse creates the current packets consisting of electron-like and hole-like excitations at the edges of the gate.
At the right edge positive charges are accelerated to the right, negative ones to the left leading to positive current pulses.
The left edge behaves the other way round.
b) At the interface between the wire and the lead an incoming current packet is transmitted with factor $1+\gamma$ and reflected with $1+\gamma$.
c) For the smooth contact, the current pulse is additionally smeared, but the total transmitted current stays the same.}
\label{fig:scattering}
\end{figure}

The setup as shown in Fig.~\ref{fig:setup} consists of a 1D helical channel with interaction parameter $K$ and mode velocity $u$ in the interval from $-L/2$ to $L/2$ connected to 1D Fermi liquid channels representing the non-interacting higher dimensional leads, so that both $u(x)$ and $K(x)$ change step-like at $\pm L/2$.
A time-dependent gate voltage is applied in the middle of the setup between $-w/2$ and $w/2$ so that $\varphi(x,t) = V_G(t) \theta(w/2-|x|)$. As shown in Fig.~\ref{fig:scattering}a), a positive voltage pulse $V_G(t) >0$ at $w/2$  creates positive current pulses travelling right and left, whereas at $-w/2$ negative current pulses travelling both directions are generated. When such a current pulse meets a step-like change of the interaction parameter from $K$ to $1$, it is partially reflected with a reflection coefficient $\gamma = \frac{1-K}{1+K}$ and transmitted with a coefficient $1+\gamma$, as shown schematically in Fig.~\ref{fig:scattering}b).
Taking into account the direction of motion, the reflection coefficient for the charge is $-\gamma$. In the long time limit, transmission and reflection of charge add up to unity, satisfying charge conservation.
For an arbitrary gate voltage pulse $V_G(t)$, we solve the equations of motion and obtain
\begin{eqnarray}
 I_R(t) = &K\frac{e^2}{h} (1+\gamma) \sum\limits^\infty_{n=0} (-\gamma)^n \left(V_G\left(t-\frac{-w+(2n+1)L}{2 u}\right) \right.
 \nonumber\\*&\left.- V_G\left(t-\frac{w+(2n+1)L}{2 u}\right)\right). \label{eq:currentpulse}
\end{eqnarray}
The gate voltage induces a current pulse $(e^2/h) K V_G$, which travels between the contacts at $\pm L/2$, as manifest in the argument of $V_G$. It is reflected multiple times with the factor $\gamma$, before it is transmitted into the lead with the factor $1+\gamma$, leading to Fabry-P\'erot behavior.\cite{Safi1995,Safi1999} In this case, $K$ can be directly extracted by comparing the ratio of successively detected pulses.

\section{Frequency-resolved transport in helical channels}
\label{SectionFreq}
\begin{figure}
\centering
\includegraphics[width=.9\columnwidth]{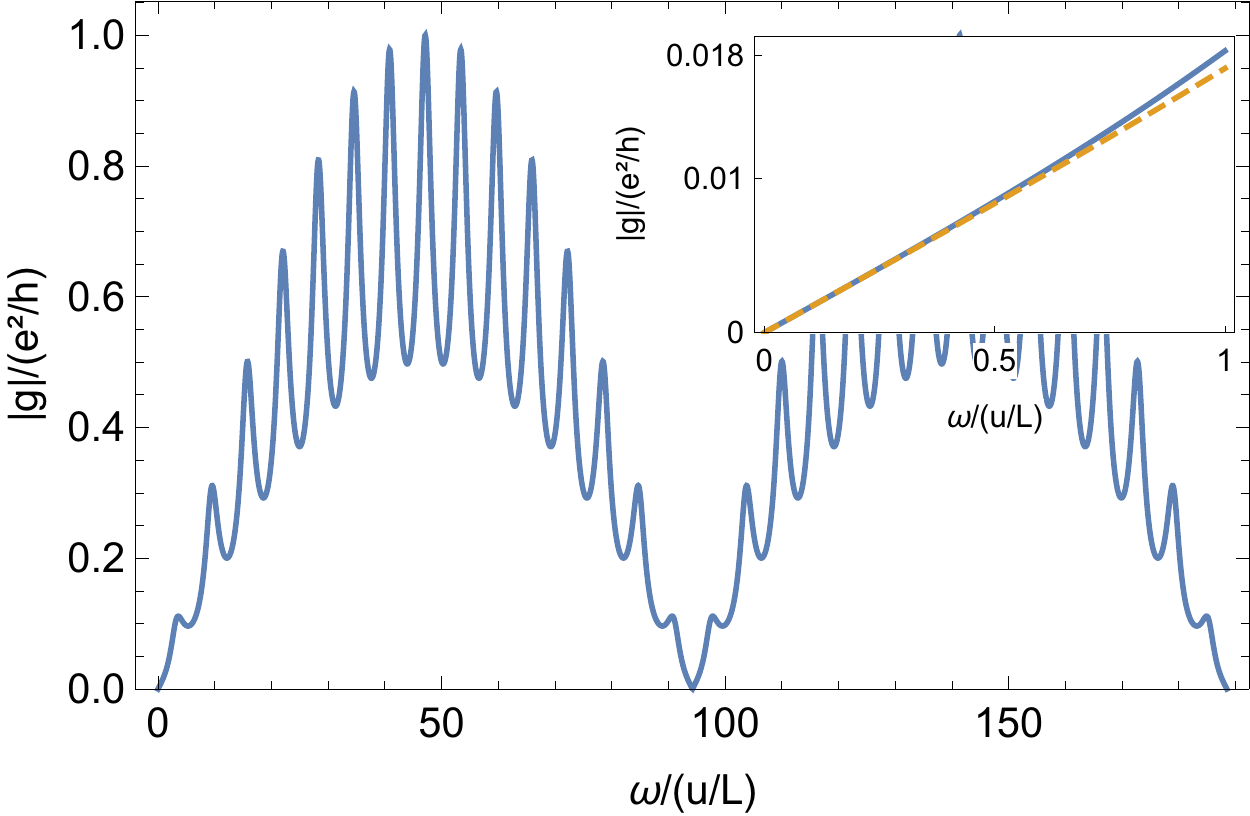}
\caption{Absolute value of the conductance $|g(\omega)|$ between gate and contact
on a one dimensional conducting channel in the Galilean case, i.e. $v_F = K u$. Here the values $K=0.5$ inside the channel and $L/w = 15$ are taken.
Inset: Comparison of $|g(\omega)|$ (solid) and its RC approximation from Eq.~\eqref{eq:Kexpansion} (dashed) in the low frequency regime.}
\label{fig:conductance}
\end{figure}
While time-resolved measurements are illustrative and can be realized with sub-nanosecond resolution, it is mostly easier to implement measurements in frequency domain. In this regard, we now consider the response to an oscillating signal $V_{G} e^{-i\omega t}$.
The linearity of Eq.~\eqref{eq:eom} results in a solely linear response, namely the conductance $g(\omega)$ such that $I_{R}(\omega)=-I_{L}(\omega)=g(\omega)V_{G}(\omega)$. Summing the series in Eq.\ref{eq:currentpulse} gives
\begin{align}
g(\omega) &= K\frac{e^2}{h} (1+\gamma) \frac{e^{i \frac{L-w}{2 u}\omega}-e^{i \frac{L+w}{2 u}\omega}}{1+\gamma e^{i\frac{L}{u}\omega}}.
\label{eq:g}
\end{align}
The conductance $g(\omega)$ is illustrated in the Fig.~\ref{fig:conductance}. It can be detected between the excitation gate and an ohmic contact, or alternatively between an excitation and a readout gate, \cite{Feve2008} both being sensitive to $I(\omega)$ in the limit of a fast-response detection. We distinguish two limits both of which enable the determination of $K$.

For low frequencies $\omega \ll u/L, u/w$, we provide a comparison with the so-called mesoscopic capacitor, a coherent series $RC$ circuit first introduced in the context of chiral edge channels of the quantum Hall effect. \cite{Buttiker1993,Gabelli2006a} To this end, we factorize the phase factor $e^{i \frac{L-w}{2 u}\omega}$ that accounts for the propagation in the edge channels on a length $\frac{L-w}{2}$ (usually not taken into account in the mesoscopic capacitor), and perform an expansion in $\omega$, which yields:
\begin{equation}
g(\omega) =e^{i \frac{L-w}{2 u}\omega}\Big(i \omega C + \omega^2 C^2R + \mathcal{O}(\omega^3)\Big),
\label{eq:Kexpansion}
\end{equation}
where
\begin{equation}
C = \frac{e^2K}{h}\frac{w}{u}, \qquad R =\frac{h}{2e^2K} \left(1-(1-K)\frac{L}{w}\right).
\label{eq:RC}
\end{equation}
We first observe that $g(\omega) \to 0$ for $\omega \to 0$, as expected for a purely capacitive gate.
In the first and second order terms, one recognizes the quantum capacitance $C$ and the charge relaxation resistance $R$.
In conventional 1D wires one can localize electrons in the quantum capacitor applying a bias voltage to a gate, which plays the role of another capacitor plate.
In the helical channel, the excitations are always delocalized due to the spin-momentum coupling.
Nevertheless, the potential well $\varphi(x)$ affects not the sign, but the absolute value of the momentum, forcing the excitations to spend more time under the gate, inducing a change of charge density by $\Delta\rho \sim \frac{e^{2}}{h} \frac{u}{K}\varphi$ [see the Eq.~\eqref{eq:eom} for the static case and subsequent discussion of the continuity].
The gate of the width $w$ collects the total charge $w\Delta\rho$, what gives the expression for the capacitance in Eq.~\eqref{eq:RC}.
When no interactions are present (i.e. $K=1$ and $\gamma=0$), these expressions reduce to $C= \frac{w}{v_F}\frac{e^2}{h}, R =\frac{h}{2e^2}$. The system is then equivalent to a non-interacting mesoscopic capacitor, as the edge states are fully decoupled, and no reflections occur at the boundaries $\pm L/2$. In particular, we recover the universal charge relaxation resistance $R=R_K/2$, where $R_K=h/e^2$ is the von Klitzing quantum of resistance. \cite{Buttiker1993,Gabelli2006a,Mora2010,Gabelli2012}
In a system with interactions, both $R$ and $C$ are modified. First, $v_F$ renormalizes to $u$, and $K$
appears in both $R$ and $C$, as already known for interacting chiral edge channels.\cite{Hamamoto2010} More importantly, $R$ exhibits a peculiar dependence on both $1-K$ and the spatial dimensions $L$ and $w$. This non-universal charge relaxation resistance is a signature of the Fabry-P\'erot behavior of the bosonic excitation. It reflects the non-chirality of helical edge channels. Through the measurement of $R$ and $C$, similarly to Ref.~\onlinecite{Gabelli2006a}, the low frequency response, gives a direct access to both the renormalized velocity $u$ and the Luttinger parameter $K$.

For higher frequencies, resonance peaks in $g(\omega)$ signal Fabry-P\'erot behavior.
As shown in Fig.~\ref{fig:conductance}, the functional dependence of $g(\omega)$ is governed by two frequencies corresponding to two different length scales, namely, the distance between the leads $L$ and the width of the gate $w$. Assuming the natural case in which $w\ll L$, one finds in the intermediate regime $\omega \approx 2\pi u/L \ll 2\pi u/w$ maxima and minima of $g$ at frequencies $\omega_{max} = (2n+1) \pi u/L$ and $\omega_{min} = 2n \pi u/L$, $n\in\mathbb{Z}$, respectively. Analysing Eq.~\eqref{eq:g}, we can see, that for any two minima and maxima in this regime, we find that
\begin{equation}
K = \frac{\omega_{min}}{\omega_{max}} \left|\frac{g(\omega_{max})}{g(\omega_{min})}\right| \; .
\label{eq:Kminmax}
\end{equation}
This formula allows for a simple determination of $K$. For larger frequencies, cancellations of $g(\omega)$ are visible and correspond to the frequency scale $\omega_{min} = 2n \pi u/w$, for which the two current contributions created on each side of the gate electrode have opposite phases and thus interfere destructively.

\section{Deviations from a step-like $K(x)$}
\label{sec_K}
\begin{figure}
\centering
\includegraphics[width=.9\columnwidth]{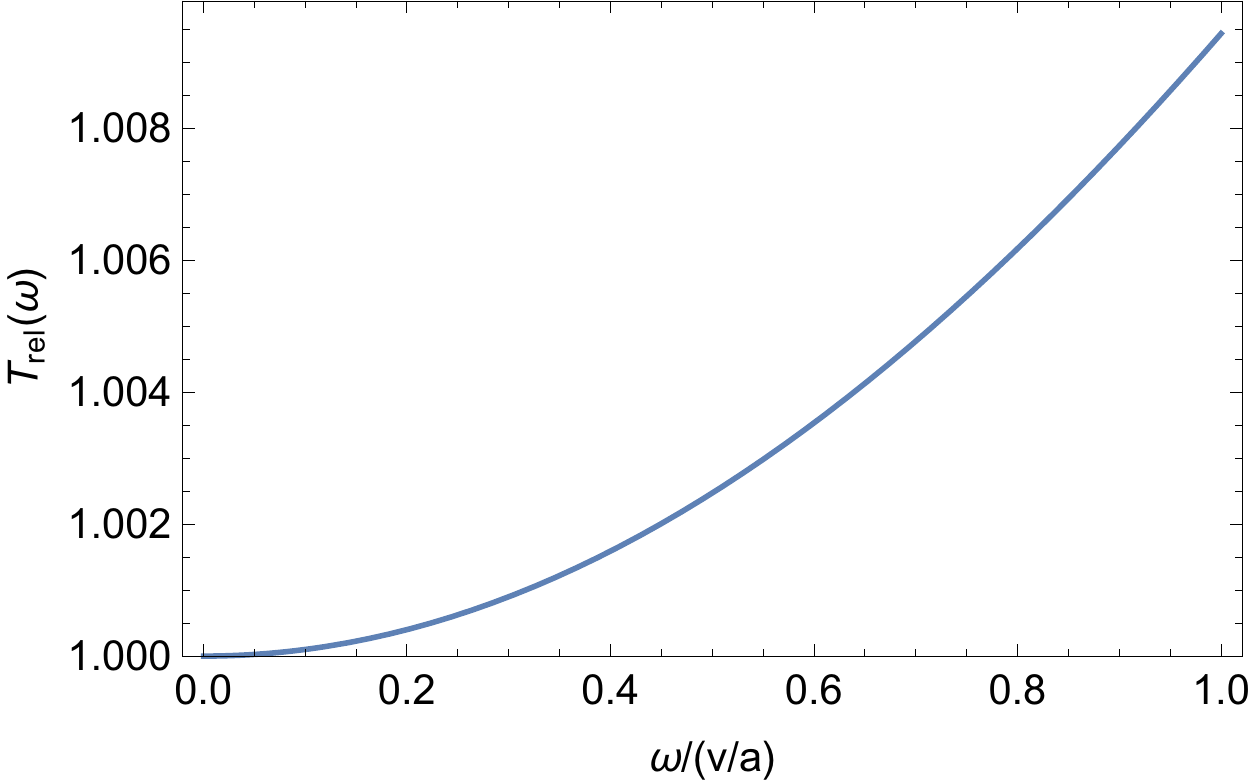}
\caption{Relative change of transmission of a sloped interaction profile at the contact between $K=0.5$ and a Luttinger liquid with $K=1$. Galilean invariance implies $u(x) = v_F/K(x)$.}
\label{fig:transmission}
\end{figure}

We consider corrections due to a non-step-like transition from an interacting $K\neq1$ region to the non-interacting leads.\cite{Thomale2011} In such a case, we argue that the frequency dependence $g(\omega)$ can also be used to probe the quality of the contacts. As we elaborated in the previous section, the transmission coefficient for a perfect step-like change of the interaction is independent of frequency. Any deviation from this idealistic model introduces a frequency dependence, as shown in Fig.~\ref{fig:transmission} for a linear spatial dependence of $K(x)$ in a Galilean invariant system. 

As the charge is still a conserved quantity and the total difference in interaction strength is fixed, the net transmitted and reflected current will stay the same regardless of the spatial variation of $K(x)$.
The form of the current pulse, however, will be modified, as schematically shown in Fig.~\ref{fig:scattering}c).
The scattering on the interface will acquire a frequency dependence, which in the limit of $\omega \to 0$ allows to recover the transmission coefficient for the step-like Luttinger parameter.
Approximating $K(x)$ by a linear dependence on an interval of a length $a$, where it changes from $K_1$ to $K_2$, we can solve Eq.~\eqref{eq:eom} analytically in the case of a Galilean invariant system for which $u(x) K(x) = v_F$.
To leading order, the frequency dependence of the transmission is then given by
\begin{equation}
 (1+\gamma) T_{rel}(\omega) \approx \frac{2}{1+K} - i \frac{2 (1-K) (2+K)}{3(1+K)^2} \frac{a \omega}{v_F}.\label{eq:Tomegaexpansion}
\end{equation}
Here, we separate the frequency dependence from the ideal transmission $1+\gamma$ using a relative transmission factor $T_{rel}(\omega)$.
The full frequency dependence and details of the calculations are given in the Appendix.
The relative change of transmission and reflection coefficients with respect to the step-like profile is shown in Fig.~\ref{fig:transmission}.
As follows from Eq.~\eqref{eq:Tomegaexpansion}, the correction to the transmission coefficient due to the soft edge of the lead is relevant at a frequency scale $\omega \gtrsim 2\pi u/a$, which is much higher than the typical frequency scales $L/u$ and $w/u$ that appear in our problem.


As the frequency dependence of $g(\omega)$ for the setup shown in Fig.~\ref{fig:setup} gets complicated in this scenario, we illustrate the effect in a simplified setup.
If the second lead is moved far away and the gate is very wide, such that $L,w \to \infty$, but $L-w$ remains finite, only the first pulse in Eq.~\eqref{eq:currentpulse} contributes to the conductance $g(\omega)$, which then takes the form:
\begin{equation}
g(\omega) = K\frac{e^2}{h} (1+\gamma) T_{rel}(\omega) e^{i \frac{L-w}{2 u}\omega}.
\end{equation}
Unlike Eq.~\eqref{eq:g}, where the strong frequency dependence is created by multiple Fabry-P\'erot-like reflection, the frequency dependence for this simplified setup is solely determined by $T_{rel}(\omega)$, because the imaginary exponent contributes to the phase only.

\section{Discussion and summary}
\label{sec_sum}

Elaborating on the experimental implementation of our proposal, the lowest characteristic frequency of the system is $\omega\sim 2\pi u/L$.
For a typical Fermi velocity of order of $u\sim1\times10^{5}\,\text{m}\,\text{s}^{-1}$, and a device length on the order of a typical mean free path of a few $100\,\text{nm}$, this characteristic frequency is of order of $1\,\text{THz}$, which is experimentally not easily accessible. Hence, long mean free paths are crucial for the feasibility of our proposal. In the helical edge states of 2D topological insulators, such as Hg(Cd)Te quantum wells, the increased mean free path (around $10\,\mu\text{m}$) allows for a substantial decrease of the characteristic frequency, down to experimentally accessible $10\,\text{GHz}$.
Our setup also allows for implementing a capacitive probe, {\it cf.} in Fig.~\ref{fig:setup}, instead of a direct measurement via ohmic leads. Capacitive coupling may be favorable for the high frequency transport measurements, as it offers a well-defined coupling capacitance with decreasing impedance for higher frequencies.

From a theoretical methodological point of view, the equation of motion perspective~\cite{Maslov1995,Thomale2011} on a helical inhomogeneous Luttinger liquid offers a promising angle to address manifold questions of DC and AC transport. For example, the equations of motion can generically contain further terms such as dissipative contributions, that are not as convenient to include at the Hamiltonian level.  

In summary, we have proposed a detailed scheme to measure the Luttinger liquid interaction parameter $K$ in the helical edge states of a 2D topological insulator. The application of a RF signal to a capacitive gate results in a $K$-dependent current response. The corresponding signal can be either measured through ohmic leads or a second capacitive gate as a probe. We propose experimental signatures in all frequency ranges. Interestingly, even in the low-frequency regime, valuable information can be extracted from such a time-of-flight experiment due to its relation to the quantum capacitance and the charge relaxation resistance.

\begin{acknowledgments}
We thank G. F\`eve, C. Mora, and A. Seidel for fruitful discussions. This work was supported by the European Research Council through ERC-StG-336012-TOPOLECTRICS. Further financial support by the DFG (SPP1666 and SFB1170 ``ToCoTronics''), the Helmholtz Foundation (VITI), and the ENB Graduate school on ``Topological Insulators'' is gratefully acknowledged. This research was supported in part by the National Science Foundation under Grant No. NSF PHY-1125915.

\end{acknowledgments}

\appendix
\section{Transmission of a sloped interaction parameter region}
Here, we would like to give some details on the calculation of transmission through a non step-like interaction profile $K(x)$ in a Galilean invariant system, i.e. the mode velocity is given by the Fermi velocity and the interaction profile as $u(x) = v_F/K(x)$.
In particular, we want to consider a linear slope in interaction strength from $K_1$ to $K_2$ over a length $a$, i.e.
\begin{equation}
 K(x) = \begin{cases}
         K_1 & x<0\\
         (K_2-K_1) x/a + K_1 & 0\leq x \leq a \\
         K_2 & x>a.
        \end{cases}\label{eq:Kslope}
\end{equation}
Assuming $\Phi(x,t) = \Phi(x,\omega) e^{-i\omega t}$, one can then separate the time dependence in Eq.~\ref{eq:eom}.
For the sloped region $0 \leq x \leq a$, the equation of motion, Eq.~\ref{eq:eom}, becomes now a Bessel differential equation
\begin{equation}
 \partial_{y}^2 \Phi(y) -\frac{2}{y} \partial_{y}\Phi(y) +\frac{\omega^2}{v_F^2} \frac{(K_2-K_1)^2}{a^2} y^2 \Phi(y) = 0, \label{eq:besseleom}
\end{equation}
where we introduced $y = x+K_1 a/(K_2-K_1)$. Left and right of this region the parameters are constant and Eq.~\ref{eq:eom} can be solved by means of plane waves, leading to the solution
\begin{equation}
 \Phi(x) = \begin{cases}
         A_1 e^{i \frac{\omega K_1}{v_F}x} + B_1 e^{-i \frac{\omega K_1}{v_F} x} & x<0\\
         y^{3/2} \left(A_2 J_{-3/4}\left(\frac{(K_2-K_1) \omega}{2 a v_F} y^2\right) \right.\\\left.+ B_2 J_{3/4}\left(\frac{(K_2-K_1) \omega}{2 a v_F} y^2\right)\right)& 0\leq x \leq a \\
         A_3 e^{i \frac{\omega K_1}{v_F}x} + B_3 e^{-i \frac{\omega K_1}{v_F} x} & x>a.
        \end{cases}\label{eq:fullPhi}
\end{equation}
We want to consider a scattering problem, where no left moving part of the field is incoming from the right, so we set $B_ 3 = 0$.
Imposing the continuity of $\Phi(x, \omega)$ and $\frac{u(x)}{K(x)} \partial_{x}\Phi(x, \omega) + e \varphi(x)$, we can express $B_1$, $B_2$, $A_2$ and $A_3$ in terms of $A_1$.
The relative change in transmission $T_{rel} = T/T_{step}$ with respect to the step-like case $T_{step} = 2K_2/(K_1+K_2)$ is then given by
%
%
%
\begin{multline}
 T_{rel} =-\frac{2 i \sqrt{2}}{\pi} \sqrt{K_1 K_2} (K_2^2-K_1^2) \times\\\times
\Biggl\{
 K_1^2 K_2^2 \frac{a\omega}{v_{F}} \Biggl[
\left(
  I_{-\frac{1}{4},-\frac{7}{4}}
+ I_{ \frac{3}{4},-\frac{3}{4}}
- I_{-\frac{7}{4},-\frac{1}{4}}
- I_{-\frac{3}{4}, \frac{3}{4}}
\right)
+\\+
i \left(
  I_{ \frac{3}{4},-\frac{7}{4}}
+ I_{-\frac{1}{4},-\frac{3}{4}}
- I_{-\frac{7}{4}, \frac{3}{4}}
+ I_{-\frac{3}{4},-\frac{1}{4}}
\right)
\Biggr]
+\\+
3 K_1^2 (K_1-K_2) \left( I_{-\frac{1}{4},-\frac{3}{4}} - i I_{ \frac{3}{4},-\frac{3}{4}} \right)
+\\+
3 K_2^2 (K_1-K_2) \left( I_{-\frac{3}{4},-\frac{1}{4}} - i I_{-\frac{3}{4}, \frac{3}{4}} \right)
\Biggr\}^{-1}
\label{eq:Tomegaanalytical}
\end{multline}
%
where $I_{a,b}=J_{a}(c_1 a\omega/v_{F})J_{b}(c_2 a\omega/v_{F})$ and $c_{1/2} = \frac{-K_{1/2}}{2(K_1-K_2)}$.
The expansion of Eq.~\ref{eq:Tomegaanalytical} for small $\omega$ leads to Eq.~\ref{eq:Tomegaexpansion} in the main text.

\bibliography{helical_ifr}

\end{document}